# Argon milling induced decoherence mechanisms in superconducting quantum circuits

J. Van Damme[1,2], Ts. Ivanov[1], P. Favia[1], T. Conard[1], J. Verjauw[1,3], R. Acharya[1,2], D. Perez Lozano[1], B. Raes[4], J. Van de Vondel[4], A. M. Vadiraj[1], M. Mongillo[1], D. Wan[1], J. De Boeck[1,2], A. Potočnik[1*], K. De Greve[1,2]

[1]*Imec, Kapeldreef 75, Leuven, B-3001, Belgium*

[2]*Department of Electrical Engineering (ESAT), KU Leuven, Leuven, B-3000, Belgium*

[3]*Department of Materials Engineering (MTM), KU Leuven, Leuven, B-3000, Belgium*

[4]*Department of Physics and Astronomy, KU Leuven, Leuven, B-3000, Belgium*

The fabrication of superconducting circuits requires multiple deposition, etch and cleaning steps, each possibly introducing material property changes and microscopic defects. In this work, we specifically investigate the process of argon milling, a potentially coherence limiting step, using niobium and aluminum superconducting resonators as a proxy for surface-limited behavior of qubits. We find that niobium microwave resonators exhibit an order of magnitude decrease in quality-factors after surface argon milling, while aluminum resonators are resilient to the same process. Extensive analysis of the niobium surface shows no change in the suboxide composition due to argon milling, while two-tone spectroscopy measurements reveal an increase in two-level system electrical dipole moments, indicating a structurally altered niobium oxide hosting larger two-level system defects. However, a short dry etch can fully recover the argon milling induced losses on niobium, offering a potential route towards state-of-the-art overlap Josephson junction qubits with niobium circuitry.

## I. INTRODUCTION

Superconducting quantum circuits show great promise for the realization of quantum information processors. The technology offers design flexibility with a variety of structures like capacitors, inductors, resonators, and Josephson junctions (JJ) that can be combined into complex systems. However, there are still multiple hurdles ahead on the road towards further upscaling and useful applications, with material imperfections as one of the principal challenges to address [1,2]. Superconducting quantum devices suffer from undesired relaxation and decoherence through interactions with two-level systems (TLS) formed by defects residing in amorphous materials, oxides, and interfaces surrounding the superconductors [3,4]. Limited qubit coherence impacts gate fidelities and requires more extensive error correction schemes. Additionally, these same TLS are responsible for random temporal fluctuations in qubit parameters, requiring active feedback and tuning [5–7]. In large superconducting quantum processors, such frequent recalibrations of the system will lead to an enormous overhead. It is therefore paramount to better understand the origins of TLS defects and develop appropriate fabrication processes with suitable materials to minimize them.

Decades of experimental research into the elusive TLS defects has revealed their predominant location at outer surfaces and interfaces of the devices [8–11]. Advances in substrate cleaning prior to metal deposition have successfully reduced losses at the substrate-metal interface [12–14], while surface oxide removal experiments [15,16] have demonstrated the importance of substrate-air and metal-air interfaces. Another important consideration is the impact of fabrication processes on the creation of defects in different materials. Ion milling is a fabrication process of interest, often used to improve ohmic contacts between two metal layers, or during the fabrication of a JJ with an overlap process [17–19]. Previous works have shown that substrate cleaning with ion milling prior to metal deposition degrades the metal-substrate interface [12,14,20,21]. Global surface ion milling in high vacuum, or $NH_3$ passivated environments [22] can increase resonator losses and reduce qubit lifetimes, indicating that the metal-air and/or substrate-air interfaces can be negatively altered by ion milling. However, the underlying nature of the milling induced losses is not entirely clear. Additionally, the effect of an oxidized ion milled superconductor, which is unavoidable for the fabrication of overlap JJ [17–19], has not been adequately studied.

All state-of-the-art qubits are currently fabricated with Al/AlOx/Al JJ [23–25]. However, other circuitry like capacitor pads, feedlines, and resonators are often fabricated from other superconducting materials in favor of higher superconducting critical temperature $T_c$ (like Nb [14]), or low-loss, self-limiting native oxide properties (like TiN [26], or Ta [23]). In this work, we characterize argon milling induced substrate-air and metal-air interface losses on two of the most frequently used superconducting materials, niobium and aluminum. Since superconducting qubits and resonators share the same surface loss mechanism, resonators are often used as a 'short-loop' proxy to study surface- and interface losses in qubits [20,27]. We fabricate superconducting resonators and subject them to different argon milling conditions. We characterize and compare their microwave loss mechanisms and analyze corresponding material changes with scanning transmission electron microscopy (STEM), energy dispersive X-ray analysis (EDS), electron energy-loss spectroscopy (EELS), atomic force microscopy (AFM), X-ray photoelectron spectroscopy (XPS), and temperature dependencies. We find that aluminum devices are not affected significantly by the argon milling conditions, while niobium devices exhibit an order of magnitude increase in microwave loss. We provide evidence for the creation of TLS with increased dipole moments inside the argon damaged niobium oxide.

## II. EXPERIMENT DESCRIPTION

In this study we conduct material loss investigation using planar, lumped element resonators (LER) inside a 3D superconducting cavity. In contrast to more widely used coplanar waveguide resonators [28,29], LER provide the possibility to characterize individual chips that undergo multiple iterations of surface treatments, since no adhesion or wire bonding is required [15]. Each chip holds six LER with different designs (Res 1 – 6, and 3.9 GHz – 5.5 GHz) placed in an aluminum cavity ($TE_{101}$ mode at 8 GHz) with adjustable input/output (I/O) ports for tuning the cavity-LER coupling. More information on the device designs can be found in Appendix B and in [15]. The resonance frequencies and quality factors of LERs are extracted by fitting a generalized Lorentzian model to the transmission scattering parameters $S_{21}$ measured at 10 mK with a vector network analyzer (VNA). All reported fitting results are acquired with the python lmfit package [30]. The experimental setup and the fitting model used for cavity coupled resonators are described in Appendix A and B, respectively.

---

*Corresponding author: anton.potocnik@imec.be

The LER devices used in this study were fabricated on 300 mm wafers in the foundry-standard cleanroom at Imec [15]. High-resistivity (>3 kΩcm) silicon substrates are cleaned with hydrofluoric (HF) acid prior to sputtering of 100 nm superconducting metal (either Nb or Al). The devices are patterned with optical lithography using a $SiO_2$ hard mask and a chlorine based dry-etch. After resist and hard mask removal, the wafers are coated with protective resist and diced (stripped with acetone afterwards). A two-minute oxygen ashing step is used to remove organic residues and to saturate oxide growth on all outer surfaces. This is the reference point for the argon milling study (reference sample). Additional samples are further exposed to a chip wide physical sputtering by argon plasma, generated using an rf-field with a gas flow of 20 $cm^3$/min and a pressure of 4.6 mTorr (using Pfeiffer Spider 630 tool). These samples are argon milled for fifteen minutes with a split of plasma rf powers: 10 W, 50 W, or 100 W [FIG. 1 (a)]. Here, the 100 W power setting is strong enough to fully consume the surface oxides of both aluminum and niobium and it corresponds to the overlap JJ fabrication condition described in [17]. The differently milled samples are, at the end, intentionally re-oxidized with the same two-minute oxygen ashing recipe, ensuring similar thickness surface oxides for the comparison of the argon milling introduced losses with the reference. The consumption of niobium or aluminum, due to the argon milling, is limited to 20 nm and the step into the silicon increases from approximately 50 nm to 100 nm. In addition, the niobium sample that was exposed to the 100 W argon treatment was, after characterization, further dry-etched with 10 seconds $SiCl_4$ to remove the argon milled top layer, followed by two minutes of oxygen ashing. All differently treated LER samples and additional processing details are summarized in TABLE I.

TABLE I. Overview of measured samples. From left to right: sample name, the substrate/metal stack, the processing performed on the device after fabrication, and the different types of characterizations performed on these samples. Single tone (ST) saturation experiment, two-tone (TT) saturation experiment, STEM, XPS, AFM, and temperature dependence of resistance (on transport bridge device that received same processing) and resonance frequency (see Appendix C).

| Name | Sub./Met. | Processing | Performed characterization | | | | | |
|---|---|---|---|---|---|---|---|---|
| | | | ST | TT | STEM | XPS | AFM | Temp. |
| Nb Reference | Si/Nb | $O_2$ plasma (2 min) | x | x | x | x | x | x |
| Nb 10 W argon | Si/Nb | 10 W argon milling (15 min) + $O_2$ plasma (2 min) | x | x | | x | | |
| Nb 50 W argon | Si/Nb | 50 W argon milling (15 min) + $O_2$ plasma (2 min) | x | x | | x | | |
| Nb 100 W argon | Si/Nb | 100 W argon milling (15 min) + $O_2$ plasma (2 min) | x | x | x | x | x | x |
| Nb 100 W argon + etch | Si/Nb | 100 W argon milling (15 min) + $O_2$ plasma (2 min) + $SiCl_4$ etch (10 sec) + $O_2$ plasma (2 min) | x | | | | | |
| Al Reference | Si/Al | $O_2$ plasma (2 min) | x | x | | | | |
| Al 100 W argon | Si/Al | 100 W argon milling (15 min) + $O_2$ plasma (2 min) | x | x | | | | |

The microwave losses of reference LER devices are limited by the surface interfaces, as illustrated by our previous work where removal of the surface oxides with HF reduced losses by almost an order of magnitude [15]. Therefore, any effect of the argon milling treatments on the substrate-air and metal-air interfaces will translate into observable changes in microwave loss. All differently treated LER samples shown in

TABLE I are compared in terms of their microwave loss using single- and two-tone [31–33] microwave power saturation experiments at 10 mK inside a dilution refrigerator, as discussed in the following section. Certain samples are then selected for further surface analysis (indicated in TABLE I) based on their microwave losses.

## III. RESULTS

### A. Single-tone loss saturation spectroscopy

Superconducting resonators measured at millikelvin temperatures exhibit a microwave power-dependent loss due to their interaction with a bath of TLS defects [34]. All tunnelling TLS defects couple to phonon modes, while the critical defects also have electrical dipole moments and therefore couple to the electrical fields of the resonators or qubits in their proximity [4,35,36]. TLS form a loss channel that broadens the resonator linewidth, which is reflected in a reduced quality factor ($Q = f_r/\kappa$, with $\kappa$ the full-width-at-half-max linewidth and $f_r$ the resonator frequency). At elevated photon occupation in the resonator, TLS with transition energies close to LER's resonance are driven into saturation, a situation of equal population of ground and excited state, prohibiting them to absorb more energy from the resonator. For a uniform energy distribution of TLS, the quality factor dependence on photon number is described by Eq. (1), derived for non-interacting TLS and uniform electrical fields ($\phi = 1$) [34]

$$\frac{1}{Q_i} = \frac{1}{Q_{\text{TLS}}} \frac{\tanh \frac{hf_r}{2k_B T}}{\sqrt{1 + \left(\frac{\bar{n}}{n_c}\right)^\phi}} + \frac{1}{Q_r}. \qquad (1)$$

The contribution of the TLS to the total loss is captured by the parameter $1/Q_{\text{TLS}}$, $\bar{n}$ is the average photon number inside the resonator, $n_c$ is the critical photon number governing an onset of TLS saturation, $T$ is the temperature (10 mK), and all residual (non-TLS) losses are represented in the parameter $1/Q_r$. Experimentally, one often finds slower power dependence ($\phi < 1$) that is attributed to an interacting TLS model [37], or geometry dependent electrical fields [38,39]. The inverse quality factors are measured as a function of microwave power on all samples in TABLE I. Results of three selected resonators (Res 2, Res 3, and Res 5) of each chip are shown in FIG. 1 (b). The solid lines are fits with the TLS saturation model of Eq. (1) and the extracted loss parameters $1/Q_{\text{TLS}}$ and $1/Q_r$ are compared in FIG. 1 (c) between samples. Our data fits well with a $\phi = 0.44 \pm 0.08$.

A striking difference is observed between the effect of argon milling on niobium, compared to aluminum resonators. The niobium resonators show an order of magnitude increase in loss between the reference sample and the 100 W argon sample, while aluminum shows resilience to the same process. Both TLS losses ($1/Q_{\text{TLS}}$) and microwave power independent residual losses ($1/Q_r$) increase on niobium samples and scale with the milling power, illustrated in FIG.

1 (b, c). Since both aluminum and niobium resonators are patterned on nominally identical high-resistivity silicon substrates, argon milling induced damage at the substrate-air interface can be ruled out as dominant cause of increased microwave loss. Argon milling introduced losses, must therefore originate from the niobium surface. This is further corroborated by a complete recovery of Q-factors on niobium resonators where, after milling and oxidation, the top surface was removed with a 10 second dry-etch [Nb 100 W argon + etch sample in FIG. 1 (b, c)].

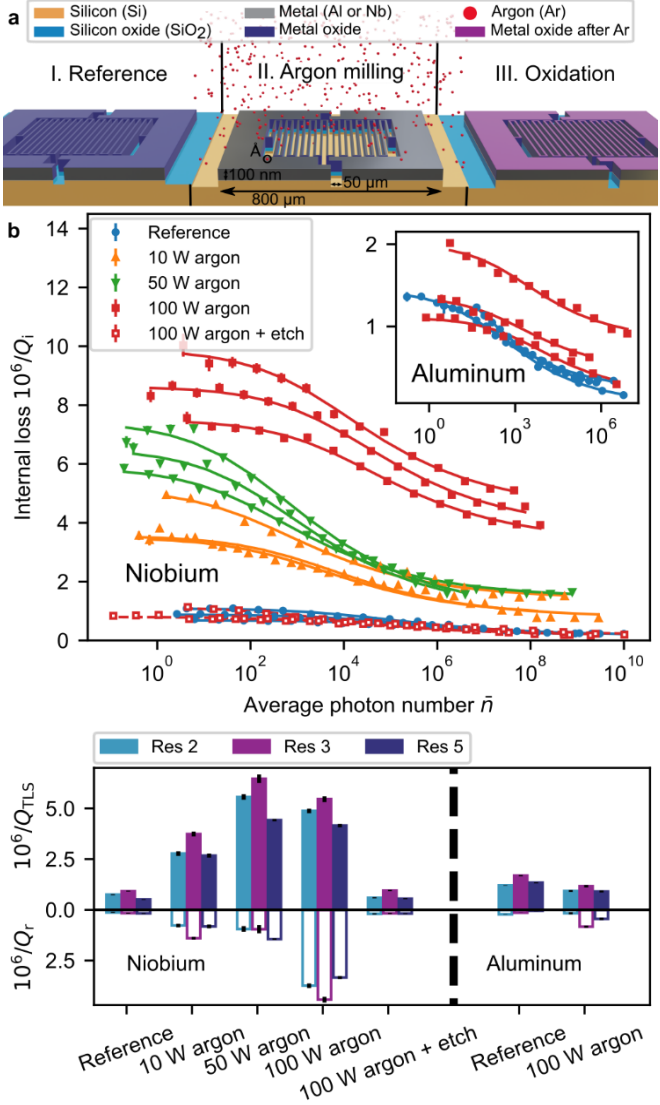

FIG. 1. (a) Schematic visualization of a cross-section of a resonator undergoing surface argon milling treatment followed by plasma oxidation. (b) Microwave power dependence of the internal losses $1/Q_i$ is compared between different samples that received varying powers of argon milling before oxidation. The same three devices (Res 2, Res 3, and Res 5) of each sample are shown. Error bars correspond to one standard deviation uncertainty extracted from the resonator line shape fitting. Continuous lines are fits with Eq. (1). The main figure contains the niobium device data, the inset shows aluminum device data. (c) Two-level system loss ($1/Q_{\text{TLS}}$) and residual loss ($1/Q_r$) extracted from the power sweeps with error bars of one standard deviation in fitting parameter uncertainty.

The niobium surface differs from the aluminum one in the complexity of the mixture of suboxides it can grow [40,41]. Aluminum oxidizes as $Al_2O_3$, while native niobium surface oxide is usually a stack of NbO, $NbO_2$, and $Nb_2O_5$, which show superconductor ($T_c \cong 1.3$ K), semiconductor (gap ∼0.1 eV) and dielectric properties, respectively [16,40,42]. The different niobium oxides can cause microwave losses through a variety of mechanisms. For example, proximity effects of a surface metallic layer (e.g. NbO) [43] or increase of quasiparticles due to magnetic impurities (e.g. paramagnetic interstitial oxygen and oxygen vacancies in $Nb_2O_5$) [44–47]. Additionally, the TLS defects hosted by the different suboxides could be drastically different. TLS in metallic glasses (like amorphous NbO) are expected to have orders of magnitude lower relaxation and coherence times due to additional scattering with electrons and would be more challenging to saturate [4]. Therefore, redistribution of the suboxides, structural reconfiguration of the suboxides and incorporation of impurities are likely causing the observed argon milling induced losses. To investigate loss sources further, we next performed surface analysis with XPS and STEM on both the Nb reference sample and the Nb 100 W argon sample.

### B. Niobium surface characterization

High angle annular dark field (HAADF) STEM images of the cross section at the niobium-air interface shows that the oxide layers prior and post argon milling are of comparable thickness (∼ 4 - 5 nm) [FIG. 2 (a, b)], indicating that the extra observed losses are intrinsic to the oxide layer and not skewed by potentially unequal amounts of lossy oxide. A striking difference is found in the appearance of layers showing different brightness in the argon milled niobium oxide [FIG. 2 (b)], indicative of a change in oxide density. The atomic concentrations extracted with EDS reveal a slight Si content (5.2 at% in the oxide layer) in the argon milled sample which was likely sputtered onto the metal surface from the nearby exposed substrate area during the milling.

To further understand the difference between samples at the niobium-air interface, we performed EELS analysis on a cross-section of the oxide layer, from which we extracted the concentration ratio O:Nb [16,48], visualized by the colormaps in FIG. 2 (c, d). Both samples show NbO suboxide (ratio O:Nb ≤ 1) at the Nb interface and an increase in oxygen content when moving towards the surface, consistent with literature reports [15,16]. The argon milled sample clearly exhibits a majority of $Nb_2O_5$ (O:Nb > 2) at the surface, while the reference sample EELS map displays a lower ratio O:Nb < 2, even at the top surface. However, surface sensitive XPS measurements [15,49] indicate a majority of $Nb_2O_5$ in all niobium samples, regardless of the argon milling conditions [FIG. 2 (f)]. The other suboxides (which are buried underneath the $Nb_2O_5$ top layer) show lower concentrations in XPS, and no clear correlation with the argon milling power could be extracted. The inconsistency with the O:Nb < 2 EELS map on the reference sample is likely due to higher surface roughness of the reference sample. Roughness along the focused ion beam (FIB) specimen thickness (∼ 40 nm) would intermix the detected signal of the thin oxide layer with the underlying niobium and the carbon on top, resulting in a reduced O:Nb ratio. This surface roughness would also explain the broadened concentration transitions at the interfaces seen in the EDS curves, compared to the sharper transitions of the argon milled sample with smoother surface [FIG. 2(a, b)].

AFM characterization of the reference and argon milled surface topography corroborate this explanation. The argon milling smoothens the niobium surface, while simultaneously increasing valley and peak areas by a factor of 5.6, approximated from the change in horizontal and vertical surface height correlation lengths [FIG. 2 (e, g)]. Surface roughness has been linked to increased loss due to enhanced participation ratio of the electric fields at the surface [50]. However, in our results, this effect is overshadowed by other loss contributions, as most losses are observed in the smoothest argon milled sample.

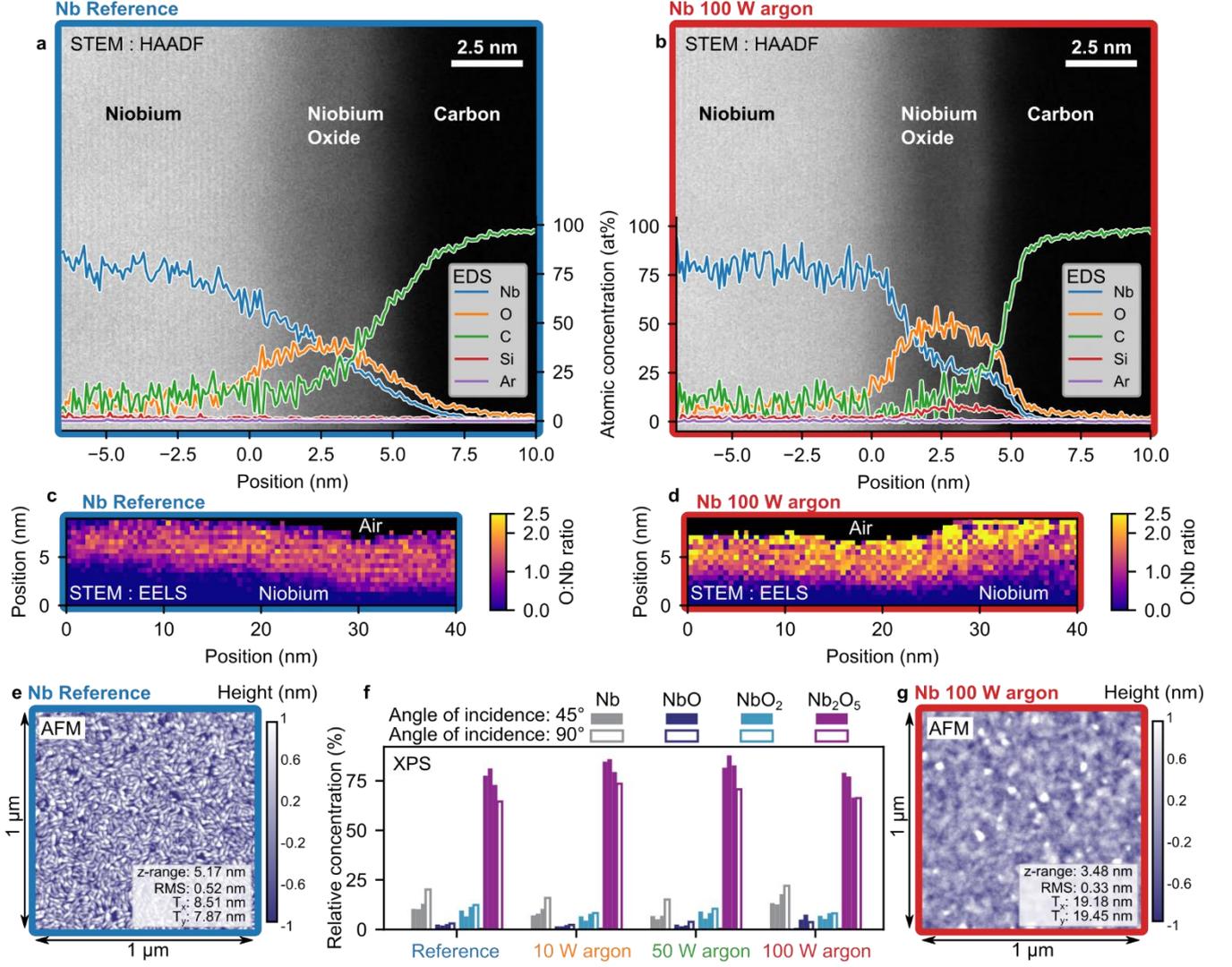

FIG. 2. Niobium-air interface analysis of a reference sample (a, c, e, f) and a sample that received the 100 W argon milling treatment (b, d, f, g). High angle annular dark field (HAADF) STEM images (a, b) are shown together with atomic concentrations of various elements as a function of the cross-sectional depth (solid lines), extracted with EDS (20 nm long region along Nb surface). EELS maps (c, d) of the niobium-air interface cross-section visualize the oxygen-to-niobium ratio throughout the oxide layer. The change in niobium surface topography due to the milling is visualized with the AFM images (e, g), which show the maximal height difference (z-range), roughness root-mean-square (RMS), horizontal, and vertical correlation lengths (Tx and Ty, respectively). Niobium suboxide concentrations for the four different argon milling conditions were extracted by fitting to the Nb3d XPS spectra (f), example shown in Appendix D. The corresponding measurement uncertainty is illustrated with multiple replicates of identically prepared samples (both resonator chips and blanket niobium chips).

Therefore, argon milling effectively smoothens the niobium surface, with little effect on the oxide layer's chemical composition apart from a slight increase in silicon content. The notable change in appearance revealed by the AFM and STEM images does indicate a change in roughness and density of the amorphous oxide layer as a consequence of niobium surface damage suffered during the milling. We revisit these findings and their possible implications in the discussion section. Additional information and details on the performed surface analysis methods can be found in Appendix D.

### C. Two-tone spectroscopy

The surface characterization results incentivize further analysis techniques sensitive to microscopic defects created by the argon milling. More insight into the TLS contribution to the losses can be gained from two-tone saturation experiments [31–33], where one signal probes the resonator lineshape, while another signal (pump) is used to saturate TLS at the frequency of the pump tone. The latter can be detuned from the resonance frequency ($f_r$), analogous to spectral hole burning [51]. Each TLS coupled to the resonator causes a complex valued frequency shift dependent on the TLS population which can be calculated by means of adiabatic elimination [31,32]:

$$\delta f_r = \sum_{j\in\{\text{TLS}\}} \frac{\Omega_{0,j}^2}{4} \frac{\langle \hat{\sigma}_z^{(j)} \rangle}{f_j - f_r + i\Gamma_2^{(j)}} \quad (2)$$

Where $hf_j$, $\Gamma_2^{(j)}$, and $\Omega_{0,j}$ are the energy, decoherence rate, and coupling rate of the $j^{\text{th}}$ TLS coupled to the resonator. The imaginary part of the complex frequency shift corresponds to added losses and the real part is a change in resonance frequency. According to Eq. (2) saturated TLS ($\langle \hat{\sigma}_z \rangle = 0$) do not add loss, nor frequency shifts to the resonator. The schematic in FIG. 3 (a) illustrates how a resonance shift can be induced with a detuned pump tone. Assuming the standard tunnelling model (STM) with a uniform energy distribution of TLS [4], a weak probe, and strong pump signal, the resonator frequency shift becomes a function of pump frequency and power as described by Eq. (3) [31–33]

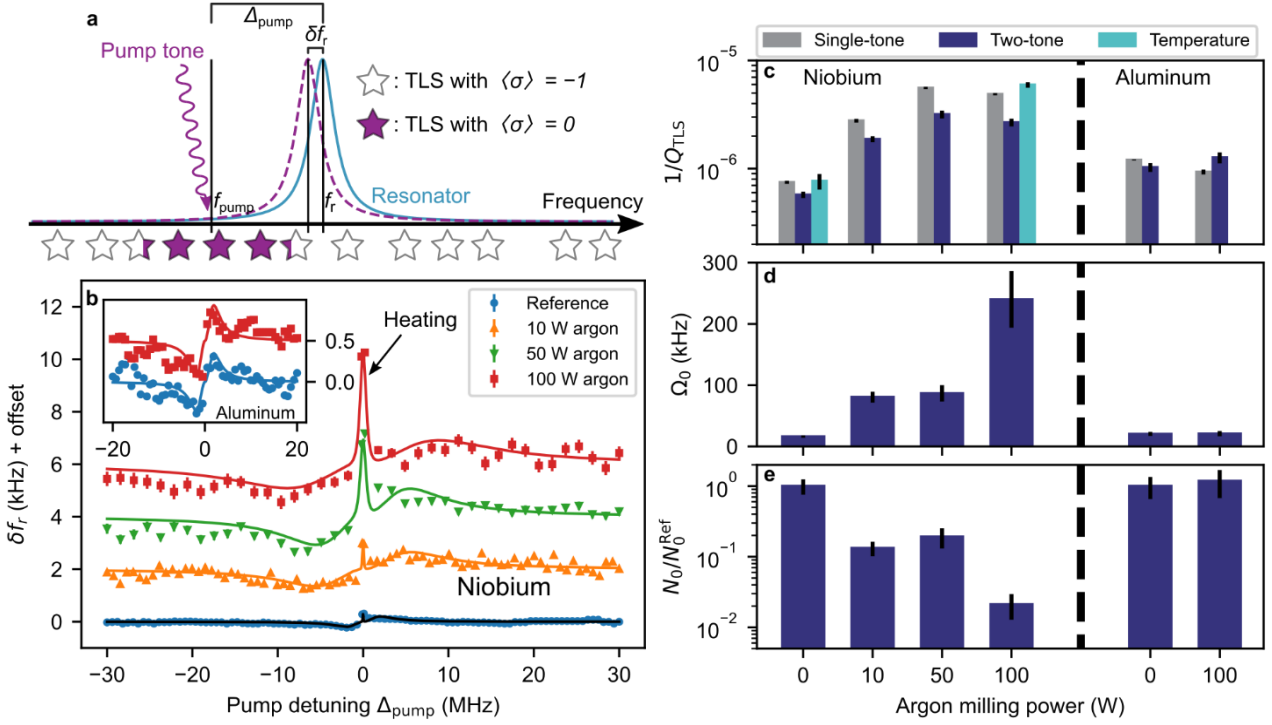

FIG. 3. Two-tone saturation measurement results. (a) Schematic of the experiment, a uniform energy distribution of two-level-system defects (stars) are partially saturated with a microwave frequency pump tone detuned with frequency Δ from the resonator. (b) The pump detuning is swept around the resonator frequency and the resulting shift in resonator frequency $\delta f_r$ is plotted. The results of Res 2 are shown for the different argon milling conditions on niobium and aluminum (inset). The data is offset on the y-axis for improved visibility. The solid lines are fits with Eq. (3) with the addition of a center peak coming from self-heating of the sample. (c) Comparison of the two-level-system loss of Res 2 on each sample extracted with the probe power sweep (FIG. 1), the pump sweep (b) and a temperature sweep (Appendix C). (d) The average two-level system coupling to the resonator expressed as the single photon Rabi-frequency extracted from the pump sweep fits. (e) The relative change in two-level system density as a function of argon milling power. All error bars are one standard deviation uncertainty in the fitting parameters.

$$\delta f_r = \frac{3\sqrt{2}}{8} \frac{f_r \tanh\left(\frac{hf_r}{k_B T}\right)}{Q_{TLS}} \frac{\Delta}{\Omega_0 \sqrt{\bar{n}}} \frac{\sqrt{1 + \frac{\Omega_0^2 \bar{n}}{2\Delta^2}} - 1}{\sqrt{1 + \frac{\Omega_0^2 \bar{n}}{2\Delta^2}} + 1}, \quad (3)$$

with experimentally tunable parameters $\Delta = f_{pump} - f_r$ the frequency detuning of the pump with the resonator, and $\bar{n}$ the average photon number inside the resonator due to the pump drive (photon number estimation is discussed in Appendix B). The resonator frequency shift depends on both the TLS loss ($1/Q_{TLS}$) and the average TLS single photon Rabi frequency ($\Omega_0$). The extrema of Eq. (3) are positioned at $\Delta = \mp \Omega_0 \sqrt{\bar{n}}/\sqrt{6}$, dependent on the coupling between resonator and TLS.

The resonance shift $\delta f_r$ as a function of pump detuning $\Delta$ is illustrated in FIG. 3 (b). At every pump frequency, the lineshape is probed with the VNA and the resonance frequency is determined. The average photon number $\bar{n}$ inside the resonator due to the pump at detuning $\Delta$ is estimated with Eq. (9). The data is then fitted with Eq. (3), from which the fitting parameters $1/Q_{TLS}$ and $\Omega_0$ are extracted.

All niobium devices show an additional frequency blue-shift in FIG. 3 (b) at small pump detuning, not captured by the model of Eq. (3). At large photon numbers (when driving close to resonance), the resonator sample locally heats up to temperatures $k_B T > hf_r$, thermally saturating the bath of TLS. Eq. (4) [34] explains the observed upwards frequency shift as function of temperature (Ψ represents the digamma function). This temperature dependence is illustrated in Appendix C where the resonance frequency is plotted as a function of temperature for the niobium reference, and 100 W argon milled samples. We have added a term accounting for heating to Eq. (3) (post fitting) to visualize it in the solid lines of FIG. 3 (b).

$$\delta f_r(T) = \frac{f_r}{\pi Q_{TLS}} \Re\left[\Psi\left(\frac{1}{2} - \frac{hf}{j2\pi k_B T}\right) - \ln\left(\frac{hf}{2\pi k_B T}\right)\right] \quad (4)$$

The extracted TLS loss component $1/Q_{TLS}$ measured for Res 2 on each sample is compared with up to three different experiments, labelled as single-tone, two-tone, and temperature respectively [FIG. 3 (c)]. The values compare reasonably well, with the observed variations attributed to a variety of factors. In the two-tone experiment, a residual TLS saturation is caused by a finite probe power ($\bar{n} \approx 10^2$). Another cause for variation is a difference in sensitivity of resonance frequency and quality factor to the TLS bath. This comes from the different dependence of the real and imaginary parts of Eq. (2) on the TLS detuning ($f_j - f_r$). In the presence of a few dominant TLS, or a non-uniform TLS energy distribution [52,53], this could result in a different $1/Q_{TLS}$ extracted from resonance frequency shifts $\delta f_r$, compared to quality-factor data.

The average TLS single photon Rabi frequency $\Omega_0$ is plotted as a function of argon milling power in FIG. 3 (d). The model of Eq. (3) is sensitive to the product $\Omega_0 \sqrt{\bar{n}}$, with $\bar{n}$ estimated up to an uncertain proportionality factor (see Appendix B) translating into uncertainty of the absolute values of $\Omega_0$. This does not change the qualitative result, nor any quantitative ratio comparisons between devices. The single photon Rabi frequency is proportional to the electrical dipole moment ($h\Omega_0 \sqrt{\bar{n}} = \mathbf{E}_{pump} \cdot \mathbf{d}$). Despite the estimated photon number uncertainty, we extract coupling rates of 20 kHz ($d/e \in [0.08; 8]$ nm as estimated in Appendix E) for our reference samples, which are comparable to literature reports [31–33,54,55]. On the niobium devices we observe an increase in dipole moments as function of argon milling power. In fact, the increase in dipole moments (up to a factor of fifteen between the Nb reference and Nb 100 W argon milled sample) exceeds the amount necessary to account for the observed loss $1/Q_{\text{TLS}} \propto d^2 N_0$, which is proportional to both the TLS density ($N_0$), and the dipole moment ($d$) squared [31]. Considering both $\Omega_0$ and $1/Q_{\text{TLS}}$ in FIG. 3 (c, d) we estimate a decrease in density of TLS with respect to the reference sample $N_0/N_0^{\text{Ref}} = Q_{\text{TLS}}^{\text{Ref}}(\Omega_0^{\text{Ref}})^2/(Q_{\text{TLS}}\Omega_0^2)$, illustrated by FIG. 3 (e), that would compensate the larger than observed losses (assuming no changes in the distribution of dipole orientations due to the milling). Furthermore, like the loss, the TLS dipole moments on the aluminum surface do not change after argon milling.

## IV. DISCUSSION

The observed resilience of the aluminum devices to the argon milling recipe could be one of the reasons why overlap JJ qubits fabricated entirely with aluminum have recently been demonstrated with lifetimes exceeding 0.1 ms [17]. Our results show that attempting a similar fabrication process with niobium would require additional care. The surface losses of argon milled niobium could be removed with an optimized over-etch during JJ top electrode patterning, except for the small contact region of this top electrode with the niobium circuitry. A similar study would have to be performed for other promising superconducting materials like tantalum, or titanium-nitride [23,24,27].

Surface analysis on niobium samples revealed no change in the niobium suboxide composition due to argon milling. However, the surface roughness is reduced by a factor of 0.63, and a layered density change is observed, with traces of Si throughout the entire oxide thickness. NbO is present at the niobium side of the metal-oxide interface in both reference and argon milled samples. We hypothesize that the additional residual losses may come from excess quasiparticles due to magnetic impurities (interstitial oxygen, and oxygen vacancies in $Nb_2O_5$ are known to have net magnetization) in the structurally altered $Nb_2O_5$ layer observed after argon milling [45,46]. Alternative hypotheses like proximity effects coming from the metallic NbO seem less likely based on the absence of significant change in critical temperature measurements between reference and argon milled niobium illustrated by FIG. 6 (a) in Appendix C. To verify our hypothesis, magnetic impurities at the niobium surface could be further investigated in future work with on-chip electron spin resonance techniques [56], point contact tunnelling spectroscopy [42], or flux temperature dependencies of SQUID loops [57].

Post argon milling TLS losses at the niobium surface increased considerably. Two-tone saturation experiments revealed a fifteen-fold increase in electrical dipole moment of the average TLS and a corresponding reduction in the density of defects with more than an order of magnitude. We speculate that on a microscopic scale the average tunnel defect changes from many single charged atoms to fewer, but larger clusters of collective atom motion [58] due to the additional argon milling surface damage. The AFM data corroborates this observation, where the surface peaks and valleys also increased in area and reduced in number. Alternatively, the presence of Si in the niobium oxide could potentially form new types of TLS with increased dipole moments. However, Si is also present after argon milling on aluminum devices, where no change in dipole moment was observed.

## V. CONCLUSION

We studied the impact of argon milling on two different superconductors, niobium, and aluminum. Niobium resonators show an order of magnitude increase in microwave loss, both the TLS and residual loss components increase as a function of argon milling power. Aluminum is resilient against the same milling conditions. XPS analysis shows no change in the niobium suboxide concentrations, STEM images reveal an altered niobium oxide with a layered structure of predominantly $Nb_2O_5$ post argon milling. Two-tone spectroscopy measurements show a fifteenfold increase in average TLS dipole moment, while AFM shows reduced roughness and enlarged topographical peak and valley areas. We speculate that the residual losses come from paramagnetic interstitial oxygen and oxygen vacancies inside the $Nb_2O_5$, and the TLS are formed by a combined motion of larger clusters of atoms. Removal of the amorphous niobium oxide layer with a short dry-etch effectively removes all argon milling induced losses, showing a path towards foundry compatible qubit fabrication with overlap JJ and niobium circuitry. Our results illustrate the importance of material and fabrication process co-optimization and the presented study method of combined loss characterization and surface analysis on superconducting resonators helped with the identification of the critical loss mechanisms in superconducting circuits.

## ACKNOWLEDGEMENTS

The authors are thankful for the imec P-line, for providing processing support, and to Danielle Vanhaeren, Olivier Richard, Laura Nelissen, Kris Paulussen, and Ilse Hoflijk for metrology support. This work is funded, in part, by the imec Industrial Affiliation Program on Quantum Computing, and by the ECSEL Joint Undertaking (JU) under grant agreement No 101007322. The JU receives support from the European Union's Horizon 2020 research and innovation program and Germany, France, Belgium, Austria, Netherlands, Finland, Israel (Please visit the project website www.matqu.eu for more information). J. VD acknowledges the support of the Research Foundation – Flanders (FWO) through the SB PhD fellowship (1S15722N). The authors would also like to thank Prof. W. De Roeck, Prof. N. de Leon, Prof. R. McDermott, Dr. J. Bejanin, and Dr. T. Maeder for insightful comments and discussions, and Prof. A. Wallraff, for providing lumped element resonator designs.

# APPENDIX A: EXPERIMENTAL SETUP

All microwave measurements reported in this work were performed inside a Bluefors LD dilution refrigerator, with the devices anchored to the 10 mK mixing chamber. Sample holders, cables, and connectors have all been carefully tested for magnetic fields and extra cryo-perm magnetic shielding around the samples was used. FIG. 4 shows a schematic of the entire measurement setup where we can switch between 6 different samples via mechanical switches connected to one input and one output line.

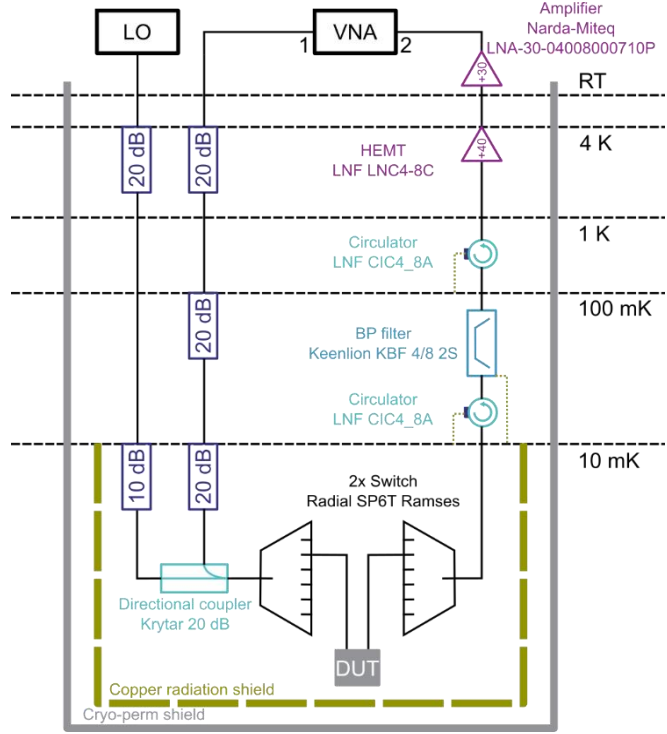

FIG. 4. Schematic of the experimental setup used in this work.

We used the Keysight P5004A streamline USB Vector Network Analyzer to probe the resonators and the Keysight M9347A local oscillator to generate the pump tone signals.

# APPENDIX B: CAVITY COUPLED RESONATOR

The samples measured in this work are silicon chips of dimension (7 mm x 4.3 mm) with 6 different planar lumped element resonator structures as shown by the microscope picture in FIG. 5 (a). The chip is placed inside an aluminum cavity without adhesive or wirebonds as illustrated by FIG. 5 (b) including two input-output (I/O) ports on the side. Non-magnetic SMA panel mount connectors can be screwed onto the cavity ports and copper spacers (not shown) can be used to tune the center pin insertion depth. The cavity lowest frequency $TE_{101}$ mode is excited and read-out via these I/O ports and is in turn coupled wirelessly to the resonator structures on the chip. Resonator designs are discussed in more detail by reference [15].

## Fitting model

The cavity coupled resonator system is described by the Hamiltonian given by Eq. (5) (including the rotating wave approximation), with $f_c = \omega_c/2\pi \approx 8$ GHz the cavity $TE_{101}$ resonance frequency, $f_r = \omega_r/2\pi \approx 5$ GHz the LER resonance frequency, and $g/2\pi \approx 50$ MHz the coupling rate between cavity and resonator. The creation and annihilation operator commutation relations are $[\hat{a}, \hat{a}^\dagger] = 1$ and $[\hat{b}, \hat{b}^\dagger] = 1$ for the cavity $TE_{101}$ mode and resonator mode respectively.

$$\hat{H} = \hbar\omega_c \hat{a}^\dagger \hat{a} + \hbar\omega_r \hat{b}^\dagger \hat{b} + \hbar g(\hat{a}\hat{b}^\dagger + \hat{a}^\dagger \hat{b}) \quad (5)$$

The coupling from the resonators on the chip to the I/O ports is assumed to be entirely mediated via the cavity mode, and any direct coupling between the small I/O port pins to the small resonator structures is neglected. The time-domain Langevin equations describing these input-output relations are given by Eq. (6), where $\kappa_A$ is the coupling rate to I/O port A, $\kappa_B$ is the coupling rate to the I/O port B, $\gamma_c$ is the internal loss rate of the cavity and $\gamma_r$ is the internal loss rate of the resonator [59].

$$\begin{cases} \dfrac{d\hat{a}}{dt} = \dfrac{-i}{\hbar}[\hat{a}, \hat{H}] - \dfrac{\kappa_A + \kappa_B + \gamma_c}{2}\hat{a} + \sqrt{\kappa_A}\hat{A}_{in} + \sqrt{\kappa_B}\hat{B}_{in} \\ \dfrac{d\hat{b}}{dt} = \dfrac{-i}{\hbar}[\hat{b}, \hat{H}] - \dfrac{\gamma_r}{2}\hat{b} \end{cases} \quad (6)$$

At ports A and B, the boundary conditions $(\hat{A}_{in} + \hat{A}_{out} = \sqrt{\kappa_A}\hat{a})$ and $(\hat{B}_{in} + \hat{B}_{out} = \sqrt{\kappa_B}\hat{b})$ hold. This set of equations can be solved using Fourier transformation and results in the following Eq. (7) for the transmission scattering parameter $S_{21}$, where we have assumed symmetrical

I/O coupling $\kappa_A = \kappa_B = \kappa$ and negligible losses of the superconducting cavity walls (high purity aluminum) $\gamma_c \ll \kappa$. We defined $\Delta_c = \omega - \omega_c$ and $\Delta_r = \omega - \omega_r$.

$$S_{21} = \left.\frac{B_{\text{out}}}{A_{\text{in}}}\right|_{B_{\text{in}}=0} = \frac{\kappa}{\kappa - i\Delta_c}\left(1 + \frac{g^2}{(\kappa - i\Delta_c)(\gamma_r/2 - i\Delta_r) + g^2}\right) \quad (7)$$

In the regime $\Delta_c \gg g, \kappa \gg \gamma_r, \Delta_r$ we can define the effective coupling rate from the perspective of the resonator as $\kappa_{\text{eff}} = \frac{g^2}{(\omega_r - \omega_c)^2}\kappa$ and approximate Eq. (7) as:

$$S_{21} \approx \frac{-i\kappa}{\Delta_c} + \frac{i\kappa_{\text{eff}}}{\omega - \widetilde{\omega}_r + i\frac{2\kappa_{\text{eff}} + \gamma_r}{2}} \quad (8)$$

$$\text{with } \widetilde{\omega}_r = \omega_r + \frac{g^2}{\omega_r - \omega_c}$$

The experimental data is measured with the VNA through the cryostat input and output lines. We account for impedance mismatches, unknown attenuation/gain and phase shifts in the fitting formula, from which we then extract the two parameters of interest $\omega_r$ and $\kappa_{\text{tot}} = 2\kappa_{\text{eff}} + \gamma_r$.

Additionally, we adjust the I/O pins for weak coupling $\kappa_{\text{eff}} \ll \gamma_r$ such that the loaded full-width-half-max (FWHM) linewidth is entirely determined by the internal losses ($\kappa_{\text{tot}} \approx \gamma_r$). Figure FIG. 5 (c) illustrates the fitting model applied to experimental data, including the distinctive "peak and dip" in the magnitude plot, a consequence of the interference between the cavity component and the resonator component. The weak coupling regime makes measurements at low powers slow and is currently a significant drawback of this type of cavity coupled resonator system, compared to hanger type CPW resonators coupled to a feedline [29].

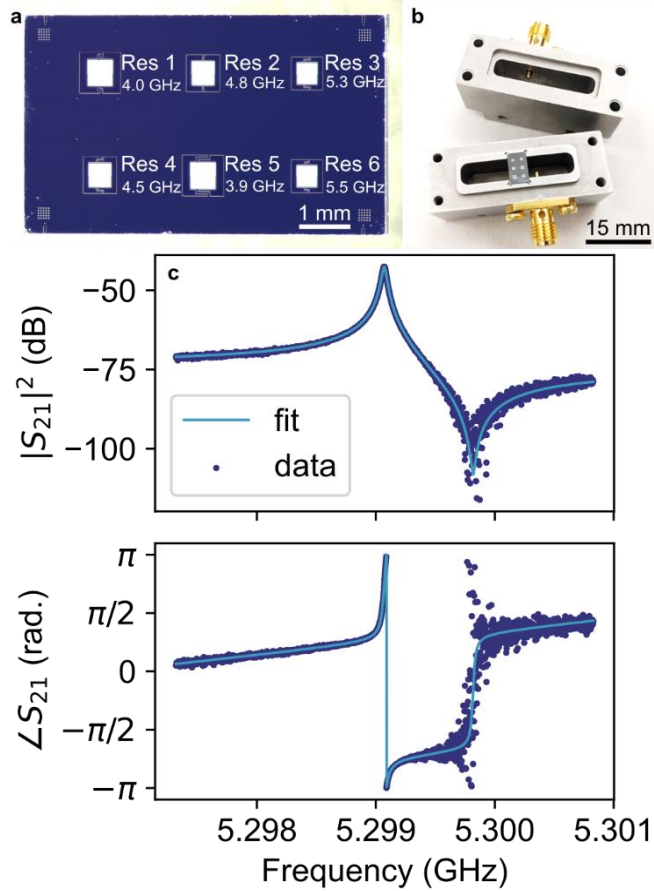

FIG. 5. Microscope picture of a silicon chip with six different superconducting lumped element resonator designs (a). The chip is loaded inside an aluminum cavity (b) with adjustable input/output pins and a $TE_{101}$ mode at 8 GHz, facilitating under-coupled read-out. (c) Example $S_{21}$ trace measured (magnitude and phase) with the VNA and fitted with Eq. (8).

## Photon number extraction

The number of photons inside the resonator due to applied input microwaves are calculated from Eq. (6) and are approximately given by Eq. (9), with $P_{\text{in}} = \langle A_{\text{in}}^\dagger A_{\text{in}}\rangle \hbar \omega_r$ the microwave power arriving at port A of the cavity.

$$\bar{n} = \langle \hat{b}^\dagger \hat{b}\rangle \approx \frac{P_{\text{in}}}{\hbar\omega_r}\frac{4\kappa_{\text{eff}}}{(2\kappa_{\text{eff}} + \gamma_r)^2 + 4(\omega - \omega_r)^2} \quad (9)$$

The photon number estimation depends on the attenuation of the input line of the cryostat ($P_{\text{in}} = G_{\text{in}} P_{\text{VNA}}$) and the coupling rate of the resonator $\kappa_{\text{eff}}$. The input attenuation of the VNA input line is estimated as $G_{\text{in}} = -94$ dB and $G_{\text{in}} = -44$ dB for the LO pump-line, based on the results of other experiments performed with superconducting qubits in the same cryostat where photon numbers could be resolved (data not included, but available upon request). The effective coupling rate to the resonators was estimated experimentally by measuring the resonator linewidths in the overcoupled case $\kappa_{\text{eff}} = \kappa g^2/\Delta^2 \gg \gamma_r$ when the I/O pins are fully inserted into the cavity. From these measurements the coupling $g$ for each resonator design was extracted and summarized in TABLE II.

TABLE II. Experimentally extracted coupling rates between the on-chip resonators and the cavity $TE_{101}$ mode as shown in FIG. 5.

| Resonator name | Res 1 | Res 2 | Res 3 | Res 4 | Res 5 | Res 6 |
|---|---|---|---|---|---|---|
| $g/2\pi$ (MHz) | 55 | 30 | 25 | 45 | 35 | 40 |

These reported values for $g$ and $G_{\text{in}}$ were used throughout this entire work and were assumed to be identical for different cooldowns and different samples (same designs and cavity dimensions). Potential errors on the photon number estimation due to incorrect input line attenuation or inaccurate coupling rate values would not change any conclusions drawn in this work.

# APPENDIX C: TEMPERATURE DEPENDENCIES

Two transport bridge structures ($L \times W \times H = 20 \times 0.2 \times 0.1\ \mu m^3$), from the same 300 mm wafer as the resonator chips measured in this work, were prepared with the reference and 100 W argon recipes. No significant impact of the milling on the superconducting properties of the niobium could be found. FIG. 6 (a) illustrates the superconducting critical temperature at 9 K for both samples at zero applied field. These measurements were repeated at different applied magnetic fields from which the Ginzburg-Landau (GL) coherence lengths $\xi(T=0)$ were extracted. The electron mean free path $l_{\text{el}}$ was calculated in the dirty limit using the BCS coherence length $\xi_0 = 38$ nm. These measurements were carried out at the Physics Department of KU Leuven and the extracted parameter values are presented in TABLE III.

TABLE III. Comparison of the superconducting parameters of thin film niobium transport bridge structures treated with the reference or 100 W argon milling post-processing recipes. We report the critical temperature ($T_c$), the GZ coherence length ($\xi(T=0)$) and the electron mean free path ($l_{\text{el}}$)

|  | Niobium reference | Niobium 100 W argon |
|---|---|---|
| $T_c$ (K) | 9.04 | 9.00 |
| $\xi(T=0)$ (nm) | 11.6 | 12.3 |
| $l_{\text{el}}$ (nm) | 4.84 | 5.44 |

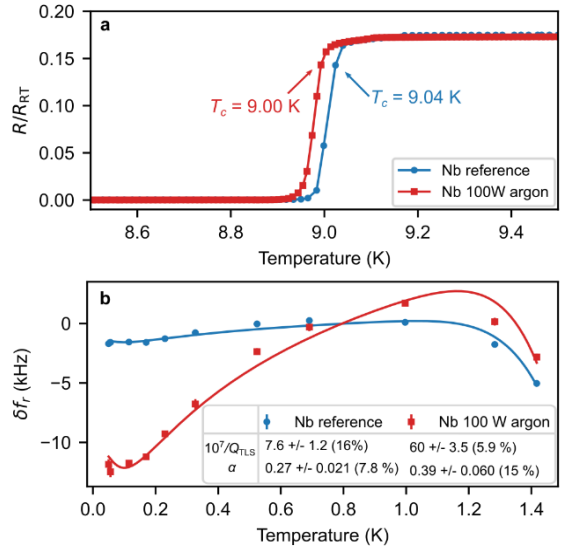

FIG. 6. Temperature dependences of reference niobium and argon milled niobium. (a) Resistance, normalized to the room temperature resistance, measured with transport bridge structures processed identically as the measured resonator chips. The critical temperature $T_c$ is extracted from the point in the superconductivity transition where the resistance is 90% of the resistance at 10 K. (b) Resonance frequency of Res 2 on each sample as function of temperature. Solid lines are fits with Eq. (10) and corresponding fitting parameters for the TLS loss ($1/Q_{\text{TLS}}$) and kinetic inductance ratio ($\alpha$) are added to the legend with their respective standard deviations.

FIG. 6 (b) compares the Res 2 resonance frequency shift as a function of temperature on the niobium reference sample and the 100 W argon milled sample. The temperature was controlled and recorded via a mixing chamber heater and temperature sensor. The resonance frequency temperature dependence is governed by TLS and quasiparticle contributions according to Eq. (10) [34].

$$\delta f_r(T) = \frac{f_r}{\pi Q_{\text{TLS}}} \Re\left[\Psi\left(\frac{1}{2} - \frac{hf}{j2\pi k_B T}\right) - \ln\left(\frac{hf}{2\pi k_B T}\right)\right] - \frac{\alpha f_r}{2}\frac{X_S(T) - X_S(0)}{X_S} \quad (10)$$

With $\Psi$ the digamma function, $\alpha$ the kinetic inductance fraction, and $X_S$ the surface inductance. In the local or dirty limit, the surface impedance is given by [34]:

$$Z_S = R_S + jX_S = \frac{j\mu_0\omega_r}{\sqrt{\frac{\omega_r l_{el}}{\sigma_n v_F \lambda_0^2}(\sigma_2(T) + j\sigma_1(T))}} \quad (11)$$

With $v_F$ the fermi velocity, $\lambda_0$ the London penetration depth, and $\sigma_1(T)/\sigma_n$ and $\sigma_2(T)/\sigma_n$ given by the Mattis-Bardeen integrals [60]. These can be evaluated into an analytical form in the regime of interest where the superconducting gap is large $\Delta_0 \gg \hbar\omega_r, k_B T$ [34]:

$$\frac{\sigma_1(T)}{\sigma_n} = \frac{4\Delta_0}{\hbar\omega}e^{-\frac{\Delta_0}{k_B T}}\sinh\left(\frac{\hbar\omega}{2k_B T}\right)K_0\left(\frac{\hbar\omega}{2k_B T}\right) \quad (12)$$

$$\frac{\sigma_2(T)}{\sigma_n} = \frac{\pi\Delta_0}{\hbar\omega}\left(1 - \sqrt{\frac{2\pi k_B T}{\Delta_0}}e^{-\frac{\Delta_0}{k_B T}} - 2e^{-\frac{\Delta_0}{k_B T}}e^{\frac{-\hbar\omega}{2k_B T}}I_0\left(\frac{\hbar\omega}{2k_B T}\right)\right) \quad (13)$$

With $I_0$ and $K_0$ the 0$^{\text{th}}$ order modified Bessel functions of the first and second kind respectively.

The fits in FIG. 6 (b) are made using Eq. (10) - (13) with free fitting parameters $1/Q_{\text{TLS}}$ and $\alpha$.

## APPENDIX D: SURFACE ANALYSIS

## XPS

Multiple XPS measurements were carried out using an Ulvac - Phi VersaProbe III instrument and an Ulvac – Phi QUANTES instrument at Imec. On the QUANTES tool, the measurements were performed under an angle of 45° (more surface sensitive) and 90° (more bulk sensitive), using a monochromatized photon beam of 1486.6 eV and a spot size of 50 microns. With the VersaProbe III instrument the analysis was done with the same 1486.6 eV photon beam under a 45° angle, a spot size of 100 microns, and a surface cleaning with a gas cluster ion beam (GCIB) sputter gun. We performed XPS on the niobium surface of samples (resonator chips and blanket niobium chips) prepared with the previously described split of argon milling powers (reference, 10 W, 50 W, and 100 W).

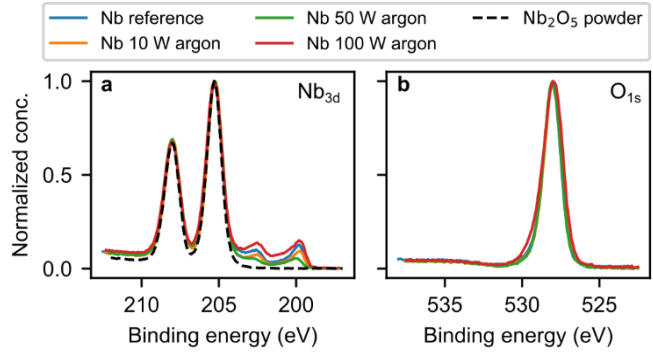

FIG. 7. XPS spectra at the Nb$_{3d}$ and O$_{1s}$ binding energies. Peak fitting with mixed Gaussian-Lorentzian peaks, adjustable mixing percentage, width, and intensity was done on this experimental data to extract the relative suboxide concentrations reported in the main text FIG. 2 (f). The data presented corresponds to blanket niobium samples (subjected to the split of argon milling conditions) and a high purity (>99.99%) Nb$_2$O$_5$ powder for tool calibration.

We attempted to find evidence for interstitial oxygen in the O$_{1s}$ peak, where this could potentially show up as side peaks in the XPS spectrum. However, after careful surface cleaning with GCIB for the removal of any adsorbed carbon and hydrogen (whose oxides typically have different binding energies), no significant changes in the O$_{1s}$ peak as function of argon milling conditions could be observed.

## STEM

The specimens were capped with carbon at room temperature for FIB lift-out extraction in Helios 450Hp. The images were taken with the Titan G2 60-300 operated at 200 kV.

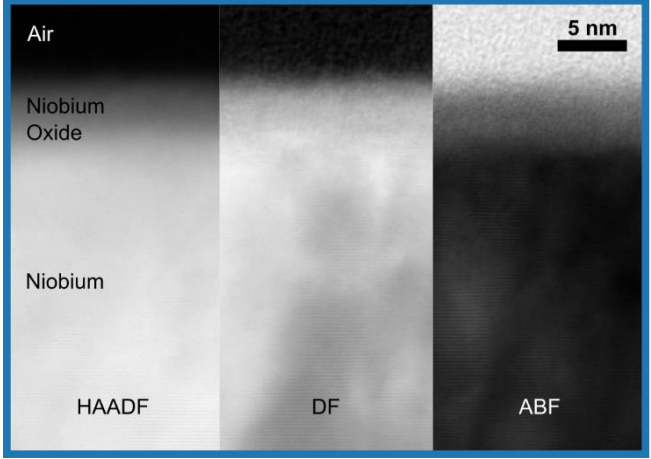 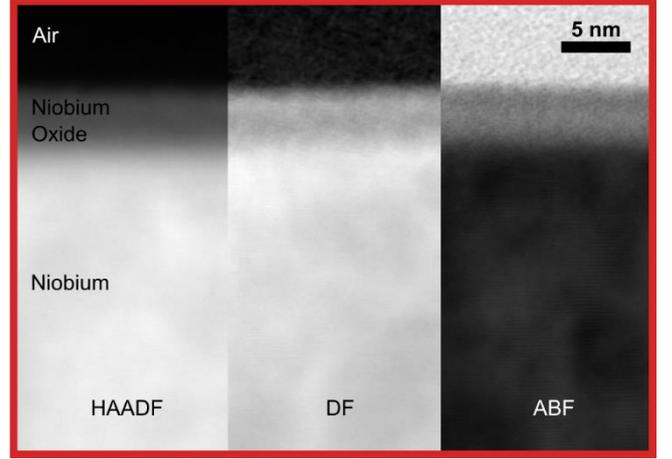

FIG. 8. STEM images of a cross-section at the niobium-air interface for a reference sample (a) and a sample subjected to the previously described 100 W argon milling recipe (b). Scattered electrons are detected with three modes: high angle annular dark field (HAADF), medium angle dark field (DF), and low angle annular bright field (ABF).

The brightness of HAADF images is to first order proportional to $\langle Z \rangle^2$ with $Z$ the atomic number and is also linearly dependent on specimen thickness. The EDS curves FIG. 2 (a, b) show traces of silicon inside the niobium oxide after the argon milling treatment. The atomic number of Si ($Z = 14$) lies in between Nb ($Z = 41$), and O ($Z = 8$) and the Si EDS content spans the entire oxide thickness, so we don't believe it is contributing to the observed layers in the 100 W argon sample. The layers are also clearly visible in medium angle and low angle STEM images (contrast sensitive to changes in density and crystallinity), which indicates a structural difference, rather than a compositional difference between the layers seen in the oxide. The surface roughness (along the thickness of the FIB sample) also plays a role in the visibility of such thin layers. Images taken on a rougher region of a 100 W argon milled sample did no longer show the presented layered oxide (data not included, but available upon request).

## AFM

The AFM measurements shown in FIG. 2 (e, g) were taken on the niobium surface of the resonator structures with the ICON PT tool using a OMCL-AC160TS-R3 probe.

## APPENDIX E: TLS DIPOLE MOMENT ESTIMATION

We numerically simulated the electric field distribution of the Res 2 design using the eigenmode solver of Ansys HFSS with a total field energy of one resonant photon $hf_r \approx 3.18 \times 10^{-24}$ J. The extracted field strength ranges across the device from 0.01 V/m to 1 V/m (FIG. 9). We can estimate the average TLS dipole moment using these field values as $d = h\Omega_0/E_{(\bar{n}=1)}$ as tabulated in TABLE IV. Despite the uncertainties in the photon number estimation, and crude approximations, we extract reasonable values for the TLS dipole moments [55,61,62].

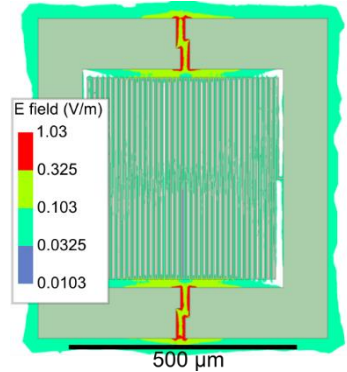

FIG. 9. Simulated electric field distribution of Res 2 for a total energy of one resonant photon at 4.8 GHz.

TABLE IV. Single photon Rabi frequencies ($\Omega_0$) of the average TLS defect for each sample measured with the two-tone spectroscopy technique described in the main text (FIG. 3). The error values correspond to one standard deviation uncertainty in the extracted fitting parameter $\Omega_0$.

|  | Al reference | Al 100 W argon | Nb reference | Nb 10 W argon | Nb 50 W argon | Nb 100 W argon |
|---|---|---|---|---|---|---|
| $\Omega_0$ (kHz) | $20.4 \pm 3.4$ | $20.9 \pm 4.3$ | $16.2 \pm 1.9$ | $80.4 \pm 9.0$ | $86.8 \pm 13$ | $240 \pm 46$ |
| †$d/e$ (nm) | 0.84 | 0.86 | 0.67 | 3.3 | 3.6 | 9.9 |

† For the electric field value $E = 0.1$ V/m (see FIG. 9).